%% file: main.tex
\documentclass[sigconf]{acmart}
\usepackage{balance}  
\usepackage{booktabs} 
\usepackage{cleveref}
\usepackage{url}

\usepackage[colorinlistoftodos,prependcaption,textsize=tiny]{todonotes}

\title{Learned Cardinalities:\\Estimating Correlated Joins with Deep Learning}

\author{Andreas Kipf}
\affiliation{\institution{Technical University of Munich}}
\email{kipf@in.tum.de}

\author{Thomas Kipf}
\affiliation{\institution{University of Amsterdam}}
\email{t.n.kipf@uva.nl}

\author{Bernhard Radke}
\affiliation{\institution{Technical University of Munich}}
\email{radke@in.tum.de}

\author{Viktor Leis}
\orcid{0000-0001-5676-8017}
\affiliation{\institution{Technical University of Munich}}
\email{leis@in.tum.de}

\author{Peter Boncz}
\affiliation{\institution{Centrum Wiskunde \& Informatica}}
\email{boncz@cwi.nl}

\author{Alfons Kemper}
\affiliation{\institution{Technical University of Munich}}
\email{kemper@in.tum.de}

\renewcommand{\shorttitle}{Learned Cardinalities}
\renewcommand{\shortauthors}{Kipf et al.}

\setcopyright{cidr19}

\acmConference[CIDR'19]{9th Biennial Conference on Innovative Data Systems Research (CIDR `19)}{January 13-16, 2019}{Asilomar, California, USA}
\acmYear{2019}
\copyrightyear{2019}
\acmISBN{}
\acmDOI{}

\begin{document}

\input{abstract}

\maketitle

\input{introduction}
\input{relatedwork}
\input{learnedcardinalities}
\input{evaluation}
\input{discussion}
\input{conclusions}
\input{acknowledgements}

\bibliographystyle{abbrv}
\bibliography{main}

\end{document}

%% file: abstract.tex
\begin{abstract}
We describe a new deep learning approach to cardinality estimation.
MSCN is a multi-set convolutional network, tailored to representing relational query plans, that employs set semantics to capture query features and true cardinalities.
MSCN builds on sampling-based estimation, addressing its weaknesses when no sampled tuples qualify a predicate, and in capturing join-crossing correlations.
Our evaluation of MSCN using a real-world dataset shows that deep learning significantly enhances the quality of cardinality estimation, which is the core problem in query optimization.
\end{abstract}

%% file: introduction.tex
\section{Introduction}
\label{sec:introduction}

Query optimization is fundamentally based on cardinality estimation.
To be able to choose between different plan alternatives, the query optimizer must have reasonably good estimates for intermediate result sizes.
It is well known, however, that the estimates produced by all widely-used database systems are routinely wrong by orders of magnitude---causing slow queries and unpredictable performance.
The biggest challenge in cardinality estimation are join-crossing correlations~\cite{qoleis,leis2018query}.
For example, in the Internet Movie Database (IMDb), French actors are more likely to participate in romantic movies than actors of other nationalities.

The question of how to better deal with this is an open area of research.
One state-of-the-art proposal in this area is Index-Based Join Sampling (IBJS)~\cite{DBLP:conf/cidr/LeisRGK017} that addresses this problem by probing qualifying base table samples against existing index structures.
However, like other sampling-based techniques, IBJS fails when there are no qualifying samples to start with (i.e., under selective base table predicates) or when no suitable indexes are available.
In such cases, these techniques usually fall back to an ``educated'' guess---causing large estimation errors.

The past decade has seen the widespread adoption of machine learning (ML), and specifically neural networks (deep learning), in many different applications and systems.
The database community also has started to explore how machine learning can be leveraged within data management systems.
Recent research therefore investigates ML for classical database problems like parameter tuning~\cite{DBLP:conf/sigmod/AkenPGZ17}, query optimization~\cite{DBLP:conf/sigmod/MarcusP18,DBLP:conf/sigmod/OrtizBGK18,joinsdeep18}, and even indexing~\cite{DBLP:conf/sigmod/KraskaBCDP18}.

We argue that machine learning is a highly promising technique for solving the cardinality estimation problem.
Estimation can be formulated as a supervised learning problem, with the input being query features and the output being the estimated cardinality.
In contrast to other problems where machine learning has been proposed like index structures~\cite{DBLP:conf/sigmod/KraskaBCDP18} and join ordering~\cite{DBLP:conf/sigmod/MarcusP18}, the current techniques based on basic per-table statistics are not very good.
In other words, an estimator based on machine learning does not have to be perfect, it just needs to be better than the current, inaccurate baseline.
Furthermore, the estimates produced by a machine learning model can directly be leveraged by existing, sophisticated enumeration algorithms and cost models without requiring any other changes to the database system.

In this paper, we propose a deep learning-based approach that learns to predict (join-crossing) correlations in the data and addresses the aforementioned weak spot of sampling-based techniques.
Our approach is based on a specialized deep learning model called multi-set convolutional network (MSCN) allowing us to express query features using sets (e.g., both $(A \Join B) \Join C$ and $A \Join (B \Join C)$ are represented as $\{A, B, C\}$).
Thus, our model does not waste any capacity for memorizing different permutations (all having the same cardinality but different costs) of a query's features, which results in smaller models and better predictions.
The join enumeration and cost model are purposely left to the query optimizer.

We evaluate our approach using the real-world IMDb dataset~\cite{qoleis} and show that our technique is more robust than sampling-based techniques and even is competitive in the sweet spot of these techniques (i.e., when there are many qualifying samples).
This is achieved using a (configurable) low footprint size of about 3\,MiB (whereas the sampling-based techniques have access to indexes covering the entire database).
These results are highly promising and indicate that ML might indeed be the right hammer for the decades-old cardinality estimation job.

%% file: relatedwork.tex
\section{Related Work}
\label{sec:relatedwork}

Deep learning has been applied to query optimization by three recent papers~\cite{DBLP:conf/sigmod/MarcusP18,DBLP:conf/sigmod/OrtizBGK18,joinsdeep18} that formulate join ordering as a {\em reinforcement learning} problem and use ML to find {\em query plans}.
This work, in contrast, applies {\em supervised learning} to solve {\em cardinality estimation} in isolation.
This focus is motivated by the fact that modern join enumeration algorithms can find the optimal join order for queries with dozens of relations~\cite{DBLP:conf/sigmod/NeumannR18}.
Cardinality estimation, on the other hand, has been called the ``Achilles heel'' of query optimization~\cite{lohmanblog} and causes most of its performance issues~\cite{qoleis}.

Twenty years ago the first approaches to use neural networks for cardinality estimation where published for UDF predicates~\cite{DBLP:conf/vldb/LakshmiZ98}.
Also, regression-based models have been used before for cardinality estimation~\cite{DBLP:conf/icde/AkdereCRUZ12}.
A semi-automatic alternative for explicit machine learning was presented in~\cite{DBLP:conf/cidr/MalikBC07}, where the feature space is partitioned using decision trees and for each split a different regression model was learned.
These early approaches did not use deep learning nor included features derived from statistics, such as our sample-based bitmaps, which encode exactly which sample tuples were selected (and we therefore believe to be good starting points for learning correlations).
The same holds for approaches that used machine learning to predict overall resource consumption: running time, memory footprint, I/O, network traffic~\cite{DBLP:journals/pvldb/LiKNC12,DBLP:conf/icde/GanapathiKDWFJP09}, although these models did include course-grained features (the estimated cardinality) based on statistics into the features.
Liu et al.~\cite{DBLP:conf/cascon/LiuXYCZ15} used modern ML for cardinality estimation, but did not focus on joins, which are the key estimation challenge~\cite{qoleis}.

Our approach builds on sampling-based estimation by including cardinalities or bitmaps derived from samples into the training signal.
Most sampling proposals create per-table samples/sketches and try to combine them intelligently in joins~\cite{DBLP:conf/icde/EstanN06,DBLP:conf/sigmod/WuNS16,DBLP:journals/pvldb/VengerovMZC15,DBLP:conf/sigmod/ChenY17}.
While these approaches work well for single-table queries, they do not capture join-crossing correlations and are vulnerable to the 0-tuple problem (cf.~Section~\ref{sec:0-tuple}).
Recent work by M{\"{u}}ller et al.~\cite{DBLP:journals/pvldb/MullerMK18} aims to reduce the 0-tuple problem for conjunctive predicates (albeit at high computational cost), but still cannot capture the basic case of a single predicate giving zero results.
Our reasonably good estimates in 0-tuple situations make MSCN improve over sampling, including even the idea of estimation on materialized join samples (join synopses~\cite{DBLP:conf/vldb/PoosalaI97}), which still would not handle 0-tuple situations.

%% file: learnedcardinalities.tex
\section{Learned Cardinalities}
\label{sec:learnedcardinalities}

From a high-level perspective, applying machine learning to the cardinality estimation problem is straightforward:
after training a supervised learning algorithm with query/output cardinality pairs, the model can be used as an estimator for other, unseen queries.
There are, however, a number of challenges that determine whether the application of machine learning will be successful:
the most important question is how to represent queries (``featurization'') and which supervised learning algorithm should be used.
Another issue is how to obtain the initial training dataset (``cold start problem'').
In the remainder of this section, we first address these questions before discussing a key idea of our approach, which is to featurize information about materialized samples.

\subsection{Set-Based Query Representation}
\label{sec:queryrepresentation}

We represent a query $q\in Q$ as a collection $(T_q,J_q,P_q)$ of a set of tables $T_q\subset T$, a set of joins $J_q\subset J$ and a set of predicates $P_q\subset P$ participating in the specific query $q$. $T$, $J$, and $P$ describe the sets of all available tables, joins, and predicates, respectively.

Each table $t\in T$ is represented by a unique \textit{one-hot} vector $v_t$ (a binary vector of length $|T|$ with a single non-zero entry, uniquely identifying a specific table) and optionally the number of qualifying base table samples or a bitmap indicating their positions. Similarly, we featurize joins $j\in J$ with a unique one-hot encoding. For predicates of the form $(col, op, val)$, we featurize columns $col$ and operators $op$ using a categorical representation with respective unique one-hot vectors, and represent $val$ as a normalized value $\in[0, 1]$, normalized using the minimum and maximum values of the respective column.

\begin{figure}[t!]
\centering
\includegraphics[width=0.75\linewidth]{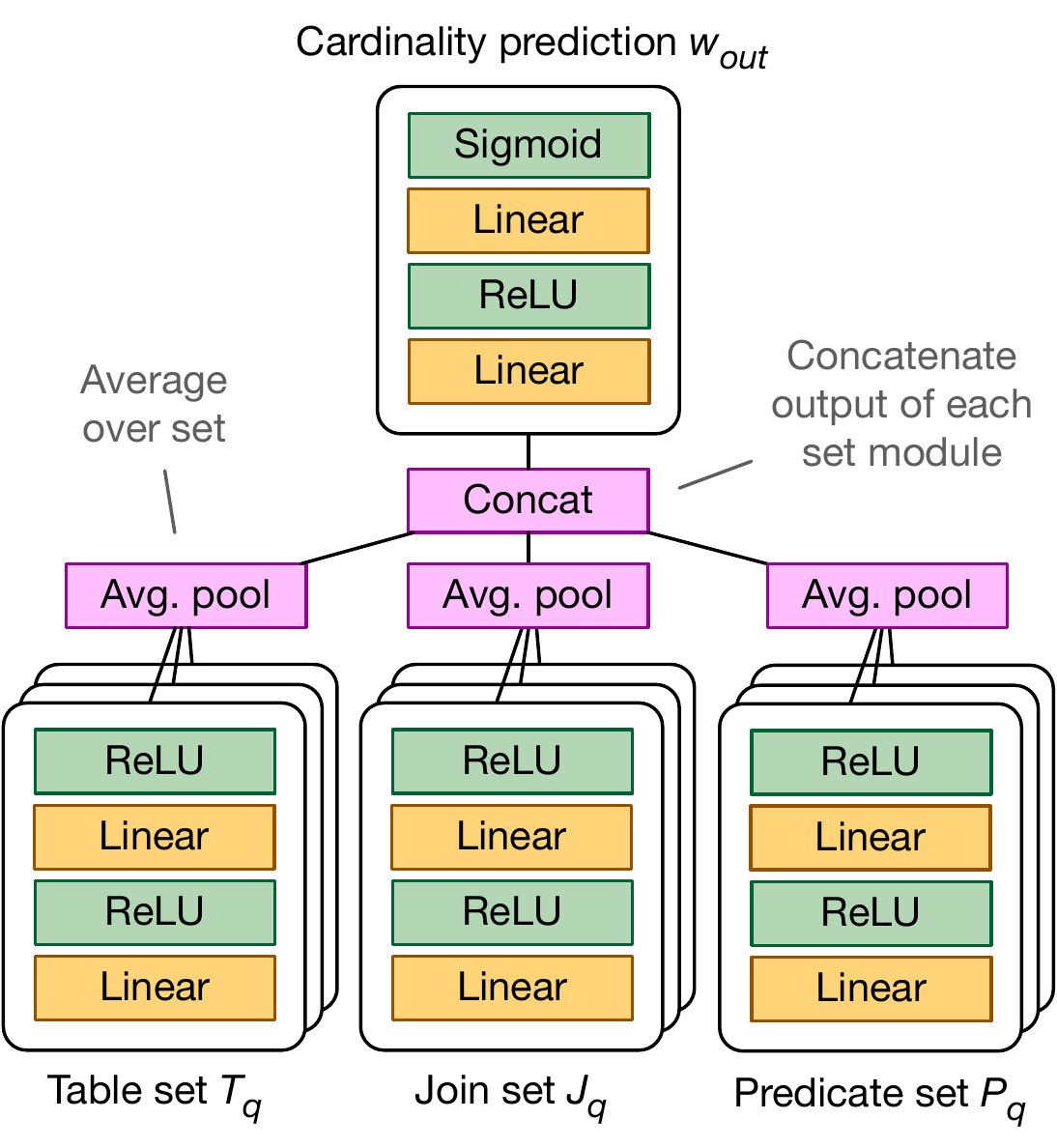}
\vspace{-0.4cm}
\caption{Architecture of our multi-set convolutional network. Tables, joins, and predicates are represented as separate modules, comprised of one two-layer neural network per set element with shared parameters. Module outputs are averaged, concatenated, and fed into a final output network.}
\label{fig:model-architecture}
\vspace{-0.4cm}
\end{figure}

Applied to the query representation $(T_q, J_q, P_q)$, our MSCN model (cf.~Figure~\ref{fig:model-architecture}) takes the following form:
\begin{align*}
\text{Table module:}\quad w_{T} &= \frac{1}{|T_q|} \textstyle\sum_{t\in T_q} \mathrm{MLP}_{T}(v_t) \\
\text{Join module:}\quad w_{J} &= \frac{1}{|J_q|} \textstyle\sum_{j\in J_q}\mathrm{MLP}_{J}(v_j) \\ 
\text{Predicate module:}\quad w_{P} &= \frac{1}{|P_q|} \textstyle\sum_{p\in P_q} \mathrm{MLP}_{P}(v_p) \\[0.4em]
\text{Merge \& predict:} \quad w_\mathrm{out} &= \mathrm{MLP}_\mathrm{out}([w_{T}, w_{J},  w_{P}])
\end{align*}

\begin{figure*}[htp!]
\centering
\includegraphics[width=0.94\linewidth]{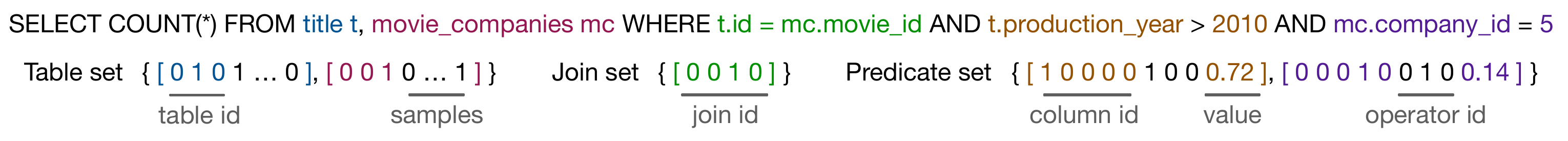}
\vspace{-0.5em}
\caption{Query featurization as sets of feature vectors.}
\label{fig:query}
\vspace{-0.4cm}
\end{figure*}

Figure~\ref{fig:query} shows an example of a featurized query.

\subsection{Model}
\label{sec:model}

Standard deep neural network architectures such as convolutional neural networks (CNNs), recurrent neural networks (RNNs), or simple multi-layer perceptrons (MLPs) are not directly applicable to this type of data structure, and would require \textit{serialization}, i.e., conversion of the data structure to an ordered sequence of elements. This poses a fundamental limitation, as the model would have to spend capacity to learn to discover the symmetries and structure of the original representation. For example, it would have to learn to discover boundaries between different sets in a data structure consisting of multiple sets of different size, and that the order of elements in the serialization of a set is arbitrary.

Given that we know the underlying structure of the data \textit{a priori}, we can bake this information into the architecture of our deep learning model and effectively provide it with an \textit{inductive bias} that facilitates generalization to unseen instances of the same structure, e.g., combinations of sets with a different number of elements not seen during training.

Here, we introduce the \textit{multi-set convolutional network} (MSCN) model. Our model architecture is inspired by recent work on \textit{Deep Sets} \cite{zaheer2017deep}, a neural network module for operating on sets. A Deep Sets module (sometimes referred to as \emph{set convolution}) rests on the observation that any function $f(S)$ on a set $S$ that is permutation invariant to the elements in $S$ can be decomposed into the form $\rho[\sum_{x\in S} \phi(x)]$ with appropriately chosen functions $\rho$ and $\phi$. For a more formal discussion and proof of this property, we refer to Zaheer et al.~\cite{zaheer2017deep}. We choose simple fully-connected multi-layer neural networks (MLPs) to parameterize the functions $\rho$ and $\phi$ and rely on their function approximation properties \cite{cybenko1989approximation} to learn flexible mappings $f(S)$ for arbitrary sets $S$. Applying a learnable mapping for each set element individually (with shared parameters) is similar to the concept of a $1\times 1$ convolution, often used in CNNs for image classification \cite{szegedy2016rethinking}.

Our query representation consists of a collection of \textit{multiple} sets, which motivates the following choice for our MSCN model architecture: for every set $S$, we learn a set-specific, per-element neural network $\mathrm{MLP}_{S}(v_s)$, i.e., applied on every feature vector $v_s$ for every element $s\in S$ individually\footnote{An alternative approach here would be to combine the feature vectors before feeding them into the MLP. For example, if there are multiple tables, each of them represented by a unique one-hot vector, we could compute the logical disjunction of these one-hot vectors and feed that into the model. Note that this approach does not work if we want to associate individual one-hot vectors with additional information such as the number of qualifying base table samples.}. The final representation $w_S$ for this set is then given by the average\footnote{Note that an average of one-hot vectors uniquely identifies the combination of one-hot vectors, e.g.~which individual tables are present in the query.} over the individual transformed representations of its elements, i.e., $w_S = 1/|S| \sum_{s\in S} \mathrm{MLP}_{S}(v_s)$. We choose an average (instead of, e.g., a simple sum) to ease generalization to different numbers of elements in the set $S$, as otherwise the overall magnitude of the signal would vary depending on the number of elements in $S$. In practice, we implement a vectorized version of our model that operates on mini-batches of data. As the number of set elements in each data sample in a mini-batch can vary, we pad all samples with zero-valued feature vectors that act as dummy set elements so that all samples within a mini-batch have the same number of set elements. We mask out dummy set elements in the averaging operation, so that only the original set elements contribute to the average.

Finally, we merge the individual set representations by concatenation and subsequently pass them through a final output MLP: $w_\mathrm{out} = \mathrm{MLP}_\mathrm{out}([w_{S_1}, w_{S_2}, \ldots, w_{S_N}])$, where $N$ is the total number of sets and $[\cdot, \cdot]$ denotes vector concatenation. Note that this representation includes the special case where each set representation $w_S$ is transformed by a subsequent individual output function (as required by the original theorem in \cite{zaheer2017deep}). One could alternatively process each $w_S$ individually first and only later merge and pass through another MLP. We decided to merge both steps into a single computation for computational efficiency.

Unless otherwise noted, all MLP modules are two-layer fully-connected neural networks with $\mathrm{ReLU}(x)=\mathrm{max}(0, x)$ activation functions. For the output MLP, we use a $\mathrm{sigmoid}(x)=1/(1+\exp(-x))$ activation function for the last layer instead and only output a scalar, so that $w_{\mathrm{out}}\in[0, 1]$. We use $\mathrm{ReLU}$ activation functions for hidden layers as they show strong empirical performance and are fast to evaluate. All other representation vectors $w_T$, $w_J$, $w_P$, and hidden layer activations of the MLPs are chosen to be vectors of dimension $d$, where $d$ is a hyperparameter, optimized on a separate validation set via grid search.

We normalize the target cardinalities $c_{\text{target}}$ as follows: we first take the logarithm to more evenly distribute target values, and then normalize to the interval $[0, 1]$ using the minimum and maximum value after logarithmization obtained from the training set\footnote{Note that this approach requires complete re-training when data changes (iff the minimum and maximum values have changed). Alternatively, one could set a high limit for the maximum value.}. The normalization is invertible, so we can recover the unnormalized cardinality from the prediction $w_{\mathrm{out}}\in[0, 1]$ of our model.

We train our model to minimize the mean \emph{q-error}~\cite{DBLP:journals/pvldb/MoerkotteNS09} $q$ ($q \geq 1$). The q-error is the factor between an estimate and the true cardinality (or vice versa). We further explored using mean-squared error and geometric mean q-error as objectives (cf.~Section~\ref{sec:optimizationmetrics}). We make use of the Adam~\cite{kingma2014adam} optimizer for training.

\subsection{Generating Training Data}
\label{sec:trainingdata}

One key challenge of all learning-based algorithms is the ``cold start problem'', i.e., how to train the model before having concrete information about the query workload.
Our approach is to obtain an initial training corpus by generating random queries based on schema information and drawing literals from actual values in the database.

A training sample consists of table identifiers, join predicates, base table predicates, and the true cardinality of the query result.
To avoid a combinatorial explosion, we only generate queries with up to two joins and let the model generalize to more joins.
Our query generator first uniformly draws the number of joins $|J_q|$ ($0\leq|J_q|\leq2$) and then uniformly selects a table that is referenced by at least one table.
For $|J_q|>0$, it then uniformly selects a new table that can join with the current set of tables (initially only one), adds the corresponding join edge to the query and (overall) repeats this process $|J_q|$ times.
For each base table $t$ in the query, it then uniformly draws the number of predicates $|P_q^{t}|$ ($0\leq|P_q^{t}|\leq\texttt{num non-key columns}$).
For each predicate, it uniformly draws the predicate type ($=$, $<$, or $>$) and selects a literal (an actual value) from the corresponding column.
We configured our query generator to only generate \emph{unique} queries.
We then execute these queries to obtain their true result cardinalities, while skipping queries with empty results.
Using this process, we obtain the initial training set for our model.

\subsection{Enriching the Training Data}
\label{sec:enriching}

A key idea of our approach is to enrich the training data with information about \emph{materialized} base table samples.
For each table in a query, we evaluate the corresponding predicates on a materialized sample and annotate the query with the \emph{number} of qualifying samples $s$ ($0\leq s \leq1000$ for 1000 materialized samples) for this table.
We perform the same steps for an (unseen) test query at estimation time allowing the ML model to utilize this knowledge.

We even take this idea one step further and annotate each table in a query with the \emph{positions} of the qualifying samples represented as bitmaps.
As we show in Section~\ref{sec:evaluation}, adding this feature has a positive impact on our join estimates since the ML model can now learn what it means if a certain sample qualifies (e.g., there might be some samples that usually have many join partners).
In other words, the model can learn to use the patterns in the bitmaps to predict output cardinalities.

\subsection{Training and Inference}
\label{sec:traininginference}

Building our model involves three steps: i) generate random (uniformly distributed) queries using schema and data information, ii) execute these queries to annotate them with their true cardinalities and information about qualifying materialized base table samples, and iii) feed this training data into an ML model.
All of these steps are performed on an immutable snapshot of the database.

To predict the cardinality of a query, the query first needs to be transformed into its feature representation (cf.~Section~\ref{sec:queryrepresentation}).
Inference itself involves a certain number of matrix multiplications, and (optionally) querying materialized base table samples (cf.~Section~\ref{sec:enriching}).
Training the model with more query samples does not increase the prediction time.
In that respect, the inference speed is largely independent from the quality of the predictions.
This is in contrast to purely sampling-based approaches that can only increase the quality of their predictions by querying more samples.

%% file: evaluation.tex
\section{Evaluation}
\label{sec:evaluation}

We evaluate our approach using the IMDb dataset which contains many correlations and therefore proves to be very challenging for cardinality estimators~\cite{qoleis}.
The dataset captures more than 2.5\,M movie titles produced over 133 years by 234,997 different companies with over 4\,M actors.

We use three different query workloads\footnote{\url{https://github.com/andreaskipf/learnedcardinalities}}: i) a \emph{synthetic} workload generated by the same query generator as our training data (using a different random seed) with 5,000 \emph{unique} queries containing both (conjunctive) equality and range predicates on non-key columns with zero to two joins, ii) another synthetic workload \emph{scale} with 500 queries designed to show how the model generalizes to more joins, and iii) \emph{JOB-light}, a workload derived from the Join Order Benchmark (JOB)~\cite{qoleis} containing 70 of the original 113 queries.
In contrast to JOB, JOB-light does not contain any predicates on strings nor disjunctions and only contains queries with one to four joins.
Most queries in JOB-light have equality predicates on dimension table attributes. The only range predicate is on \texttt{production\_year}.
Table~\ref{tab:joindistributions} shows the distribution of queries with respect to the number of joins in the three query workloads. The non-uniform distribution in the synthetic workload is caused by our elimination of duplicate queries.

\begin{table}[]
  \small
\begin{tabular}{@{}lrrrrrr@{}}
\toprule
number of joins & 0     & 1     & 2     & 3  & 4  & overall \\ \midrule
synthetic       & 1636  & 1407  & 1957  & 0  & 0  & 5000    \\
scale           & 100   & 100   & 100   & 100 & 100  & 500    \\
JOB-light       & 0     & 3     & 32    & 23 & 12 & 70      \\ \bottomrule
\end{tabular}
\caption{Distribution of joins.}
\label{tab:joindistributions}
\vspace{-1.5em}
\end{table}

As competitors we use PostgreSQL version 10.3, Random Sampling (RS), and Index-Based Join Sampling (IBJS)~\cite{DBLP:conf/cidr/LeisRGK017}.
RS executes base table predicates on materialized samples to estimate base table cardinalities and assumes independence for estimating joins. If there are no qualifying samples for a conjunctive predicate, it tries to evaluate the conjuncts individually and eventually falls back to using the number of distinct values (of the column with the most selective conjunct) to estimate the selectivity.
IBJS represents the state-of-the-art for estimating joins and probes qualifying base table samples against existing index structures.
Our IBJS implementation uses the same fallback mechanism as RS.

We train and test our model on an Amazon Web Services (AWS) ml.p2.xlarge instance using the PyTorch framework\footnote{\url{https://pytorch.org/}} and use CUDA.
We use 100,000 random queries with zero to two joins and 1,000 materialized samples as training data (cf.~Section~\ref{sec:trainingdata}).
We split the training data into 90\% training and 10\% validation samples.
To obtain true cardinalities for our training data, we use HyPer~\cite{hyper}.

\subsection{Estimation Quality}

\begin{figure}[t!]
\centering
\includegraphics[width=1.0\linewidth]{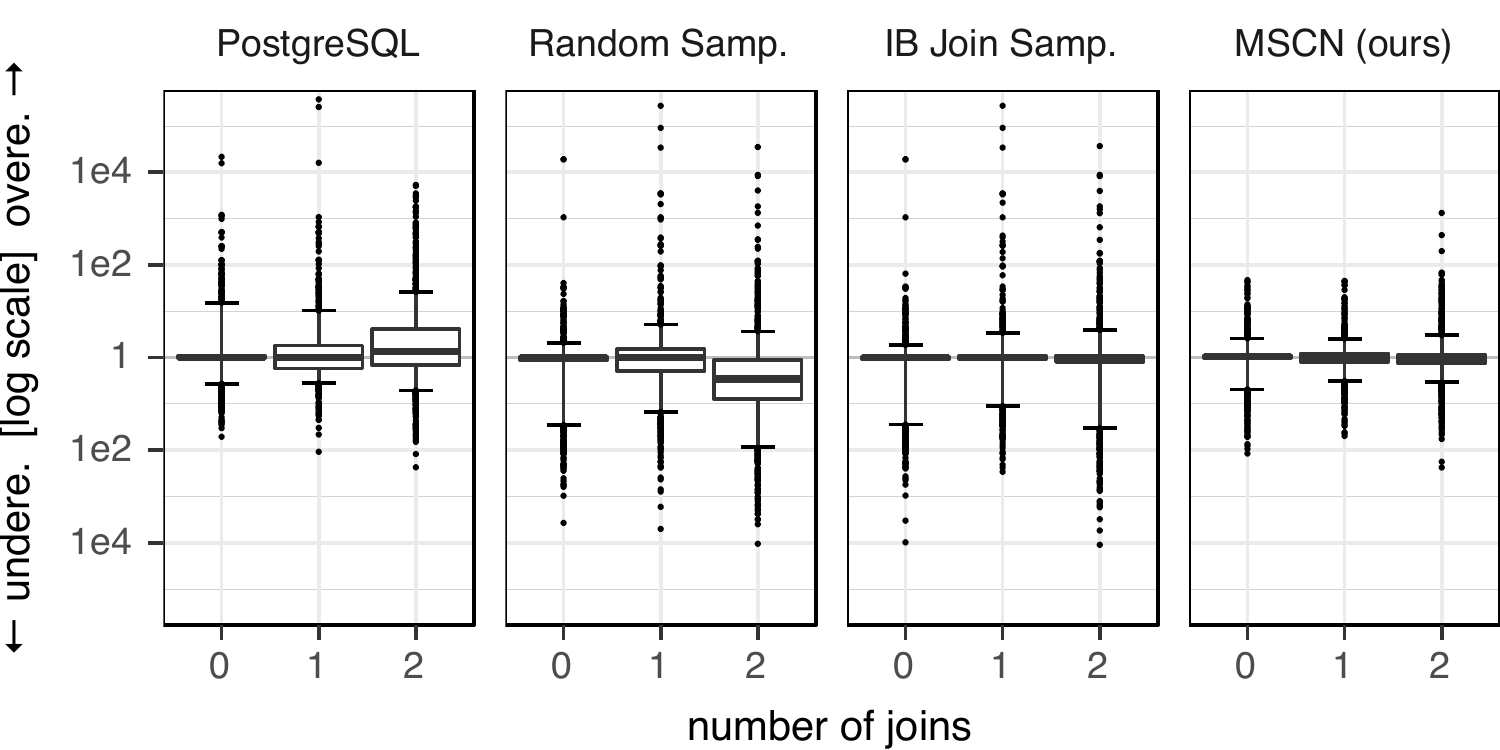}
\vspace{-0.5cm}
\caption{Estimation errors on the synthetic workload. The box boundaries are at the 25th/75th percentiles and the horizontal ``whisker'' lines mark the 95th percentiles.}
\label{fig:synthetic-qerror}
\vspace{-0.4cm}
\end{figure}

Figure~\ref{fig:synthetic-qerror} shows the q-error of MSCN compared to our competitors.
While PostgreSQL's errors are more skewed towards the positive spectrum, RS tends to underestimate joins, which stems from the fact that it assumes independence.
IBJS performs extremely well in the median and 75th percentile but (like RS) suffers from empty base table samples.
MSCN is competitive with IBJS in the median while being significantly more robust.
Considering that IBJS is using much more data---in the form of large primary and foreign key indexes---in contrast to the very small state MSCN is using (less than 3\,MiB), MSCN captures (join-crossing) correlations reasonably well and does not suffer as much from 0-tuple situations (cf.~Section~\ref{sec:0-tuple}).
To provide more details, we also show the median, percentiles, maximum, and mean q-errors in Table~\ref{tab:synthetic-qerror}.
While IBJS provides the best median estimates, MSCN outperforms the competitors by up to two orders of magnitude at the end of the distribution.

\begin{table}
  \small
\begin{tabular}{@{}lllllll@{}}
\toprule
                 & median        & 90th          & 95th          & 99th          & max           & mean          \\ \midrule
PostgreSQL       & 1.69          & 9.57          & 23.9          & 465           & 373901        & 154           \\
Random Samp.     & 1.89          & 19.2          & 53.4          & 587           & 272501        & 125           \\
IB Join Samp.    & \textbf{1.09} & 9.93          & 33.2          & 295           & 272514        & 118           \\
MSCN (ours)      & 1.18          & \textbf{3.32} & \textbf{6.84} & \textbf{30.51} & \textbf{1322} & \textbf{2.89} \\ \midrule
\end{tabular}
\caption{Estimation errors on the synthetic workload.}
\label{tab:synthetic-qerror}
\vspace{-2.0em}
\end{table}

\subsection{0-Tuple Situations}
\label{sec:0-tuple}

Purely sampling-based approaches suffer from empty base table samples (0-tuple situations) which can occur under selective predicates.
While this situation can be mitigated using, e.g., more samples or employing more sophisticated---yet still sampling-based---techniques (e.g., \cite{DBLP:journals/pvldb/MullerMK18}), it remains inherently difficult to address by these techniques.
In this experiment, we show that deep learning, and MSCN in particular, can handle such situations fairly well.

In fact, 376 (22\%) of the 1636 base table queries in the synthetic workload have empty samples (using MSCN's random seed).
We will use this subset of queries to illustrate how MSCN deals with situations where it \emph{cannot} build upon (runtime) sampling information (i.e., all bitmaps only contain zeros).
We also include Random Sampling (which uses the same random seed---i.e., the same set of materialized samples as MSCN) and PostgreSQL in this experiment.

\begin{table}[]
  \small
\begin{tabular}{@{}lllllll@{}}
\toprule
                 & median        & 90th          & 95th          & 99th          & max           & mean          \\ \midrule
PostgreSQL       & 4.78          & 62.8          & 107           & 1141          & 21522         & 133           \\
Random Samp.     & 9.13          & 80.1          & 173           & 993           & 19009         & 147           \\
MSCN             & \textbf{2.94} & \textbf{13.6} & \textbf{28.4} & \textbf{56.9} & \textbf{119} & \textbf{6.89} \\ \bottomrule
\end{tabular}
\caption{Estimation errors of 376 base table queries with empty samples in the synthetic workload.}
\label{tab:0-tuple-qerror}
\vspace{-2.0em}
\end{table}

The results, shown in Table~\ref{tab:0-tuple-qerror}, demonstrate that MSCN addresses the weak spot of purely sampling-based techniques and therefore would complement them well.

Recall that Random Sampling extrapolates the output cardinality based on the number of qualifying samples (zero in this case).
Thus, it cannot simply extrapolate from this number and has to fall back to an educated guess---in our RS implementation either using the product of selectivities of individual conjuncts or using the number of distinct values of the column with the most selective predicate.
Independent of the concrete implementation of this fallback, it remains an educated guess.
MSCN, in contrast, can use the signal of individual query features (in this case the specific table and predicate features) to provide a more precise estimate.

\subsection{Removing Model Features}

\begin{figure}[t!]
\centering
\includegraphics[width=1.0\linewidth]{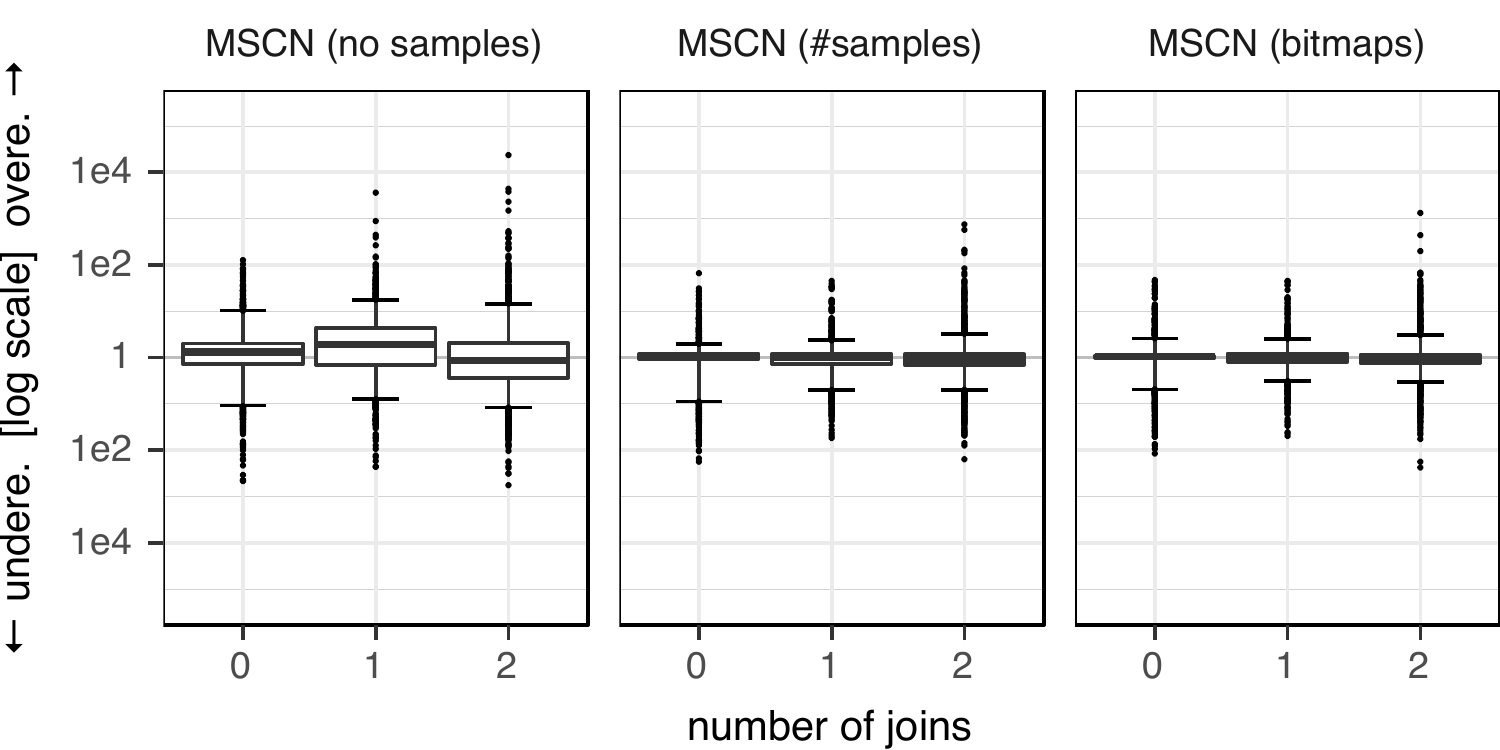}
\vspace{-0.5cm}
\caption{Estimation errors on the synthetic workload with different model variants.}
\label{fig:synthetic-qerror-ablation}
\vspace{-1.0em}
\end{figure}

Next, we highlight the contributions of individual model features to the prediction quality (cf.~Figure~\ref{fig:synthetic-qerror-ablation}).
MSCN (no samples) is the model without any (runtime) sampling features, MSCN (\#samples) represents the model with one cardinality (i.e., the number of qualifying samples) per base table, and MSCN (bitmaps) denotes the full model with one bitmap per base table.

MSCN (no samples) produces reasonable estimates with an overall 95th percentile q-error of 25.3, purely relying on (inexpensive to obtain) query features.
Adding sample cardinality information to the model improves both base table and join estimates.
The 95th percentile q-errors of base table, one join, and two join estimates reduce by 1.72$\times$, 3.60$\times$, and 3.61$\times$, respectively.
Replacing cardinalities with bitmaps further improves these numbers by 1.47$\times$, 1.35$\times$, and 1.04$\times$.
This shows that the model can use the information embedded in the bitmaps to provide better estimates.

\subsection{Generalizing to More Joins}
\label{sec:morejoins}

To estimate a larger query, one can of course break the query down into smaller sub queries, estimate them individually using the model, and combine their selectivities.
However, this means that we would need to assume independence between two sub queries which is known to deliver poor estimates with real-world datasets such as IMDb (cf.~join estimates of Random Sampling in Section~\ref{sec:evaluation}).

\begin{figure}[t!]
\centering
\includegraphics[width=1.0\linewidth]{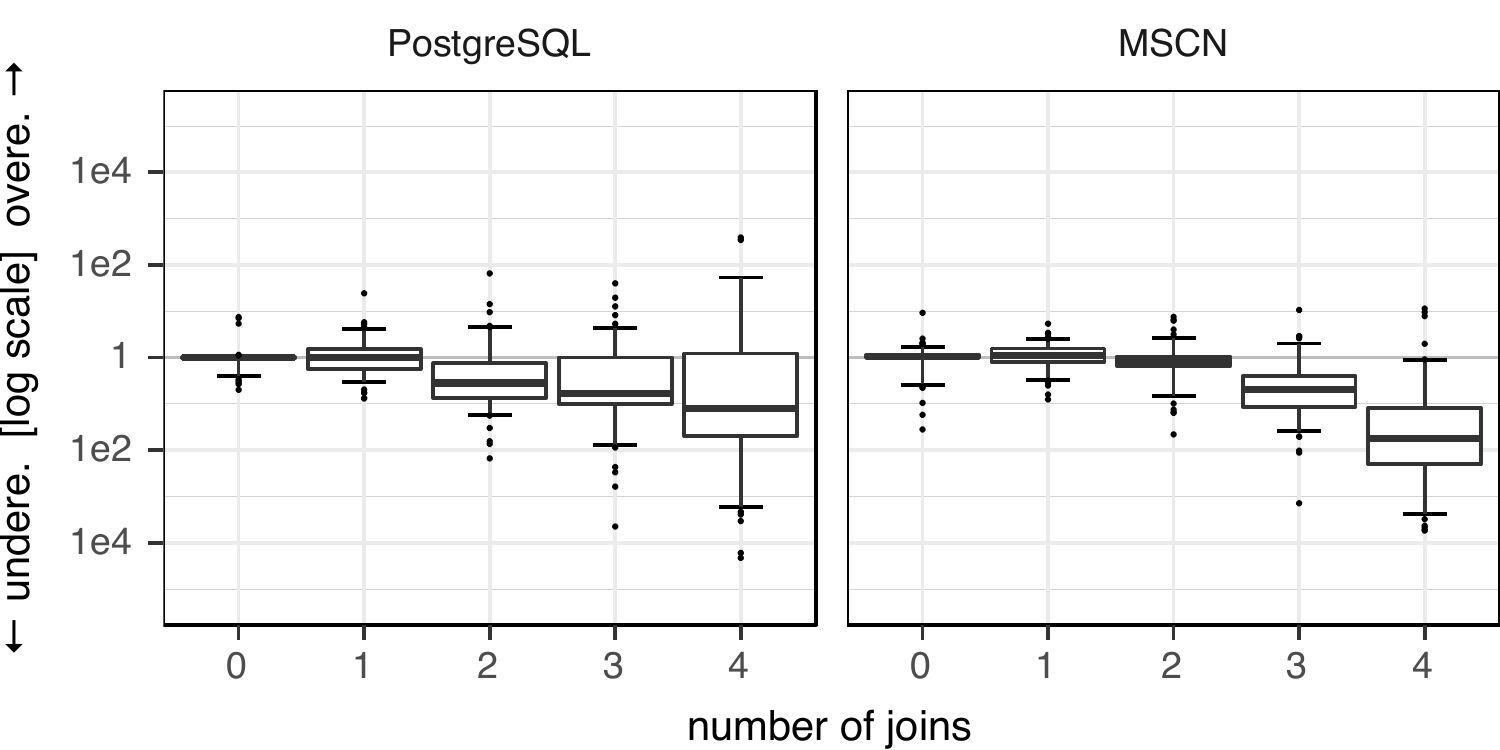}
\vspace{-0.5cm}
\caption{Estimation errors on the scale workload showing how MSCN generalizes to queries with more joins.}
\label{fig:scale-qerror}
\vspace{-1.0em}
\end{figure}

The question that we want to answer in this experiment is how MSCN can generalize to queries with more joins than it was trained on.
For this purpose, we use the \emph{scale} workload with 500 queries with zero to four joins (100 queries each).
Recall that we trained the model only with queries that have between zero and two joins.
Thus, this experiment shows how the model can estimate queries with three and four joins \emph{without} having seen such queries during training (cf.~Figure~\ref{fig:scale-qerror}).
From two to three joins, the 95th percentile q-error increases from 7.66 to 38.6.
To give a point of reference, PostgreSQL has a 95th percentile q-error of 78.0 for the same queries.
And finally, with four joins, MSCN's 95th percentile q-error increases further to 2,397 (PostgreSQL: 4,077).

Note that 58 out of the 500 queries in this workload exceed the maximum cardinality seen during training.
12 of these queries have three joins and another 46 have four joins.
When excluding these outliers, the 95th percentile q-errors for three and four joins decrease to 23.8 and 175, respectively.

\subsection{JOB-light}
\label{sec:job-light}

To show how MSCN generalizes to a workload that was not generated by our query generator, we use JOB-light.

\begin{table}
  \small
\begin{tabular}{@{}lllllll@{}}
\toprule
                 & median        & 90th          & 95th          & 99th          & max           & mean          \\ \midrule
PostgreSQL       & 7.93          & 164           & 1104          & 2912          & 3477          & 174           \\
Random Samp.     & 11.5          & 198           & 4073          & 22748         & 23992         & 1046          \\
IB Join Samp.    & \textbf{1.59} & 150           & 3198          & 14309         & 15775         & 590           \\
MSCN             & 3.82          & \textbf{78.4} & \textbf{362} & \textbf{927} & \textbf{1110} & \textbf{57.9}   \\ \bottomrule
\end{tabular}
\caption{Estimation errors on the JOB-light workload.}
\label{tab:job-qerror}
\vspace{-1.0em}
\end{table}

Table~\ref{tab:job-qerror} shows the estimation errors.
Recall that most queries in JOB-light have equality predicates on dimension table attributes.
Considering that MSCN was trained with a uniform distribution between $=$, $<$, and $>$ predicates, it performs reasonably well.
Also, JOB-light contains many queries with a closed range predicate on \texttt{production\_year}, while the training data only contains open range predicates.
Note that JOB-light also includes five queries that exceed the maximum cardinality that MSCN was trained on.
Without these queries, the 95th percentile q-error is 115.

In summary, this experiment shows that MSCN can generalize to workloads with distributions different from the training data.

\subsection{Hyperparameter Tuning}
\label{sec:hyperparameters}

We tuned the hyperparameters of our model, including the number of epochs (the number of passes over the training set), the batch size (the size of a mini-batch), the number of hidden units, and the learning rate.
More hidden units means larger model sizes and increased training and prediction costs with the upside of allowing the model to capture more data, while learning rate and batch size both influence convergence behavior during training.

We varied the number of epochs (100, 200), the batch size (64, 128, 256, 512, 1024, 2048), the number of hidden units (64, 128, 256, 512, 1024, 2048), and fixed the learning rate to 0.001, resulting in 72 different configurations.
For each configuration, we trained three models\footnote{Note that the weights of the neural network are initialized using a different random seed in each training run. To provide reasonably stable numbers, we tested each configuration three times.} using 90,000 samples and evaluated their performance on the validation set consisting of 10,000 samples.
On average over the three runs, the configuration with 100 epochs, a batch size of 1024 samples, and 256 hidden units performed best on the validation data.
Across many settings, we observed that 100 epochs perform better than 200 epochs.
This is an effect of overfitting: the model captures the noise in the training data such that it negatively impacts its prediction quality.
Overall, we found that our model performs well across a wide variety of settings.
In fact, the mean q-error only varied by 1\% within the best 10 configurations and by 21\% between the best and the worst configuration.
We also experimented with different learning rates (0.001, 0.005, 0.0001) and found 0.001 to perform best.
We thus use 100 epochs, a batch size of 1024, 256 hidden units, and a learning rate of 0.001 as our default configuration.

\subsection{Model Costs}
\label{sec:modelcosts}

\begin{figure}[t!]
\centering
\includegraphics[width=1.0\linewidth]{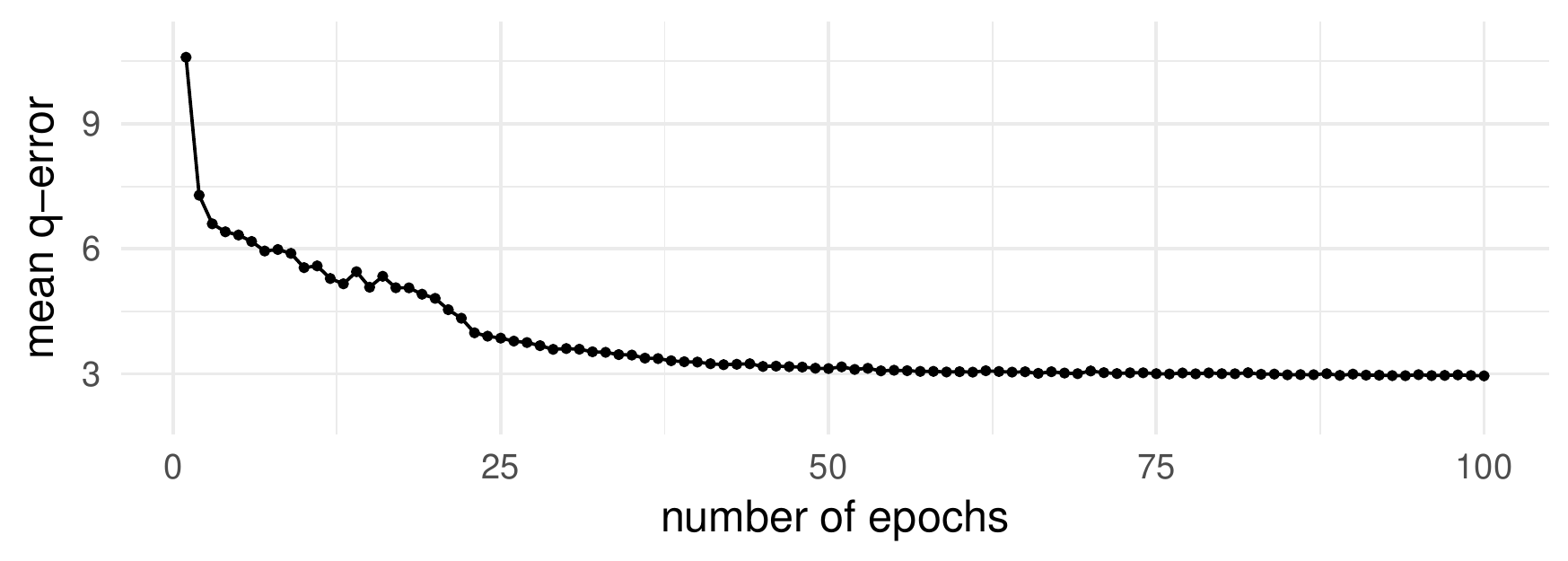}
\caption{Convergence of the mean q-error on the validation set with the number of epochs.}
\label{fig:epochs}
\vspace{-1.0em}
\end{figure}

Next, we analyze the training, inference, and space costs of MSCN with our default hyperparameters.
Figure~\ref{fig:epochs} shows how the validation set error (the mean q-error of all queries in the validation set) decreases with more epochs.
The model requires fewer than 75 passes (over the 90,000 training queries) to converge to a mean q-error of around 3 on the 10,000 validation queries.
An average training run with 100 epochs (measured across three runs) takes almost 39 minutes.

The prediction time of our model is in the order of a few milliseconds, including the overhead introduced by the PyTorch framework.
In theory (neglecting the PyTorch overhead), a prediction using a deep learning model (as stated earlier) is dominated by matrix multiplications which can be accelerated using modern GPUs.
We thus expect performance-tuned implementations of our model to achieve very low prediction latencies.
Since we incorporate sampling information, the end-to-end prediction time will be in the same order of magnitude as that of (per-table) sampling techniques.

The size of our model (when serialized to disk) is 1.6\,MiB, 1.6\,MiB, and 2.6\,MiB for MSCN (no samples), MSCN (\#samples), and MSCN (bitmaps), respectively.

\subsection{Optimization Metrics}
\label{sec:optimizationmetrics}

Besides optimizing the mean q-error, we also explored using mean-squared error and geometric mean q-error as optimization goals.
Mean-squared error would optimize the \emph{squared} differences between the predicted and true cardinalities.
Since we are more interested in minimizing the \emph{factor} between the predicted and the true cardinalities (q-error) and use this metric for our evaluation, optimizing the q-error directly yielded better results.
Optimizing the geometric mean of the q-error makes the model put less emphasis on heavy outliers (that would lead to large errors).
While this approach looked promising at first, it turned out to be not as reliable as optimizing the mean q-error.

%% file: discussion.tex
\section{Discussion}
\label{sec:discussion}

We have shown that our model can beat state-of-the-art approaches for cardinality estimation.
It does a good job in addressing 0-tuple situations and in capturing join-crossing correlations, especially when combined with runtime sampling.
To make it suitable for general-purpose cardinality estimation, it can be extended into multiple dimensions, including complex predicates, uncertainty estimation, and updatability.
In the following, we will discuss these and sketch possible solutions.

\paragraph{Generalization}
MSCN can to some extent generalize to queries with more joins than seen during training (cf.~Section~\ref{sec:morejoins}).
Nevertheless, generalizing to queries that are not in the vicinity of the training data remains challenging.

Of course, our model can be trained with queries from an actual workload or their structures.
In practice, we could replace any literals in user queries with placeholders to be filled with actual values from the database.
This would allow us to focus on the relevant joins and predicates.

\paragraph{Adaptive training}
To improve training quality, we could \emph{adaptively} generate training samples: based on the error distribution of queries in the validation set, we would generate new training samples that shine more light on difficult parts of the schema.

\paragraph{Strings}
A simple addition to our current implementation are equality predicates on strings.
To support these, we could hash the string literals to a (small) integer domain.
Then an equality predicate on a string is essentially the same as an equality predicate on an ID column where the model also needs to process a non-linear input signal.

\paragraph{Complex predicates}
Currently, our model can only estimate queries with predicate types that it has seen during training.
Complex predicates, such as LIKE or disjunctions, are not yet supported since we currently do not represent them in the model.
An idea to allow for any complex predicate would be to purely rely on the sampling bitmaps in such cases. 
Note that this would make our model vulnerable to 0-tuple situations.
To mitigate that problem, we could featurize information from histograms.
Also, the distribution of bitmap patterns might vary significantly from simple predicates observed at training time, to more complicated predicates at test time, which can make generalization challenging.

\paragraph{More bitmaps}
At the moment, we use a single bitmap indicating the qualifying samples per base table.
To increase the likelihood for qualifying samples, we could additionally use one bitmap per predicate.
For example, for a query with two conjunctive base table predicates, we would have one bitmap for each predicate, and another bitmap representing the conjunction.
In a column store that evaluates one column at a time, we can obtain this information almost for free.
We have already shown that MSCN can use the information embedded in the bitmaps to make better predictions.
We expect that it would benefit from the patterns in these additional bitmaps.

This approach should also help MSCN with estimating queries with arbitrary (complex) predicates where it needs to rely on information from the (many) bitmaps.
Of course, this approach does not work in 0-tuple situations, or more specifically in situations where none of the (predicate) bitmaps indicates any qualifying samples.

\paragraph{Uncertainty estimation}
An open question is when to actually trust the model and rely on its predictions.
One approach is to use strict constraints for generating the training data and enforce them at runtime, i.e., only use the model when all constraints hold (i.e., only PK/FK joins, only equality predicates on certain columns).
A more appealing approach would be to implement uncertainty estimation into the model.
However, for a model like ours, this is a non-trivial task and still an area of active research.
There are some recent methods~\cite{guo2017calibration,kendall2017uncertainties,lakshminarayanan2017simple} that we plan to investigate in future work.

\paragraph{Updates}
Throughout this work, we have assumed an immutable (read-only) database.
To handle data and schema changes, we can either completely re-train the model or we can apply some modifications to our model that allow for incremental training.

Complete re-training comes with considerable compute costs (for re-executing queries to obtain up-to-date cardinalities and for re-training the model) but would allow us to use a different data encoding.
For example, we could use larger one-hot vectors to accommodate for new tables and we could re-normalize values in case of new minimum or maximum values.
Queries (training samples) of which we know to still have the same cardinality (e.g., since there has not been any update to the respective data range) would of course not need to re-executed.

In contrast, incremental training (as implied by its name) would not require us to re-train with the original set of samples.
Instead, we could re-use the model state and only apply new samples.
One challenge with incremental training is to accommodate changes in the data encoding, including one-hot encodings and the normalization of values.
To recall, there are two types of values that we normalize: literals in predicates (actual column values) and output cardinalities (labels).
For both types, setting a high limit on the maximum value seems most appropriate.
The main challenge, however, is to address \emph{catastrophic forgetting}, which is an effect that can be observed with neural networks when data distribution shifts over time.
The network would overfit to the most recent data and forget what it has learned in the past.
Addressing this problem is an area of active research with some recent proposals~\cite{DBLP:journals/corr/KirkpatrickPRVD16}.

%% file: conclusions.tex
\section{Conclusions}
\label{sec:conclusions}

We have introduced a new approach to cardinality estimation based on MSCN, a new deep learning model.
We have trained MSCN with generated queries, uniformly distributed within a constrained search space.
We have shown that it can learn (join-crossing) correlations and that it addresses the weak spot of sampling-based techniques, which is when no samples qualify.
Our model is a first step towards reliable ML-based cardinality estimation and can be extended into multiple dimensions, including complex predicates, uncertainty estimation, and updatability.

Another application of our set-based model is the prediction of the number of unique values in a column or in a combination of columns (i.e., estimating the result size of a group-by operator).
This is another hard problem where current approaches achieve undesirable results and where machine learning seems promising.

%% file: acknowledgements.tex
\section{Acknowledgements}
\label{sec:acknowledgements}

This work has been partially supported by the German Federal Ministry of Education and Research (BMBF) grant \texttt{01IS12057} (FASTDATA).
It is further part of the TUM Living Lab Connected Mobility (TUM LLCM) project and has been funded by the Bavarian Ministry of Economic Affairs, Energy and Technology (StMWi) through the Center Digitisation.Bavaria, an initiative of the Bavarian State Government.
T.K. acknowledges funding by SAP SE.

\balance

%% file: main.bbl
\begin{thebibliography}{10}

\bibitem{DBLP:conf/icde/AkdereCRUZ12}
M.~Akdere, U.~{\c{C}}etintemel, M.~Riondato, E.~Upfal, and S.~B. Zdonik.
\newblock Learning-based query performance modeling and prediction.
\newblock In {\em ICDE}, pages 390--401, 2012.

\bibitem{DBLP:conf/sigmod/AkenPGZ17}
D.~V. Aken, A.~Pavlo, G.~J. Gordon, and B.~Zhang.
\newblock Automatic database management system tuning through large-scale
  machine learning.
\newblock In {\em SIGMOD}, 2017.

\bibitem{DBLP:conf/sigmod/ChenY17}
Y.~Chen and K.~Yi.
\newblock Two-level sampling for join size estimation.
\newblock In {\em SIGMOD}, 2017.

\bibitem{cybenko1989approximation}
G.~Cybenko.
\newblock Approximation by superpositions of a sigmoidal function.
\newblock {\em Mathematics of control, signals and systems}, 2(4), 1989.

\bibitem{DBLP:conf/icde/EstanN06}
C.~Estan and J.~F. Naughton.
\newblock End-biased samples for join cardinality estimation.
\newblock In {\em {ICDE}}, 2006.

\bibitem{DBLP:conf/icde/GanapathiKDWFJP09}
A.~Ganapathi, H.~A. Kuno, U.~Dayal, J.~L. Wiener, A.~Fox, M.~I. Jordan, and
  D.~A. Patterson.
\newblock Predicting multiple metrics for queries: Better decisions enabled by
  machine learning.
\newblock In {\em ICDE}, pages 592--603, 2009.

\bibitem{guo2017calibration}
C.~Guo, G.~Pleiss, Y.~Sun, and K.~Q. Weinberger.
\newblock On calibration of modern neural networks.
\newblock In {\em ICML}, 2017.

\bibitem{hyper}
A.~Kemper and T.~Neumann.
\newblock {HyPer: A Hybrid OLTP \& OLAP Main Memory Database System Based on
  Virtual Memory Snapshots}.
\newblock In {\em ICDE}, pages 195--206. {IEEE} Computer Society, Apr. 2011.

\bibitem{kendall2017uncertainties}
A.~Kendall and Y.~Gal.
\newblock What uncertainties do we need in bayesian deep learning for computer
  vision?
\newblock In {\em Advances in neural information processing systems}, pages
  5574--5584, 2017.

\bibitem{kingma2014adam}
D.~P. Kingma and J.~Ba.
\newblock Adam: A method for stochastic optimization.
\newblock {\em arXiv:1412.6980}, 2014.

\bibitem{DBLP:journals/corr/KirkpatrickPRVD16}
J.~Kirkpatrick, R.~Pascanu, N.~C. Rabinowitz, J.~Veness, G.~Desjardins, A.~A.
  Rusu, K.~Milan, J.~Quan, T.~Ramalho, A.~Grabska{-}Barwinska, D.~Hassabis,
  C.~Clopath, D.~Kumaran, and R.~Hadsell.
\newblock Overcoming catastrophic forgetting in neural networks.
\newblock {\em CoRR}, abs/1612.00796, 2016.

\bibitem{DBLP:conf/sigmod/KraskaBCDP18}
T.~Kraska, A.~Beutel, E.~H. Chi, J.~Dean, and N.~Polyzotis.
\newblock The case for learned index structures.
\newblock In {\em SIGMOD}, 2018.

\bibitem{joinsdeep18}
S.~Krishnan, Z.~Yang, K.~Goldberg, J.~Hellerstein, and I.~Stoica.
\newblock Learning to optimize join queries with deep reinforcement learning.
\newblock {\em arXiv:1808.03196}, 2018.

\bibitem{DBLP:conf/vldb/LakshmiZ98}
M.~S. Lakshmi and S.~Zhou.
\newblock Selectivity estimation in extensible databases - {A} neural network
  approach.
\newblock In {\em VLDB}, pages 623--627, 1998.

\bibitem{lakshminarayanan2017simple}
B.~Lakshminarayanan, A.~Pritzel, and C.~Blundell.
\newblock Simple and scalable predictive uncertainty estimation using deep
  ensembles.
\newblock In {\em Advances in Neural Information Processing Systems}, pages
  6402--6413, 2017.

\bibitem{qoleis}
V.~Leis, A.~Gubichev, A.~Mirchev, P.~Boncz, A.~Kemper, and T.~Neumann.
\newblock How good are query optimizers, really?
\newblock {\em PVLDB}, 9(3), 2015.

\bibitem{DBLP:conf/cidr/LeisRGK017}
V.~Leis, B.~Radke, A.~Gubichev, A.~Kemper, and T.~Neumann.
\newblock Cardinality estimation done right: Index-based join sampling.
\newblock In {\em CIDR}, 2017.

\bibitem{leis2018query}
V.~Leis, B.~Radke, A.~Gubichev, A.~Mirchev, P.~Boncz, A.~Kemper, and
  T.~Neumann.
\newblock Query optimization through the looking glass, and what we found
  running the {Join Order Benchmark}.
\newblock {\em The VLDB Journal}, 2018.

\bibitem{DBLP:journals/pvldb/LiKNC12}
J.~Li, A.~C. K{\"{o}}nig, V.~R. Narasayya, and S.~Chaudhuri.
\newblock Robust estimation of resource consumption for {SQL} queries using
  statistical techniques.
\newblock {\em {PVLDB}}, 5(11):1555--1566, 2012.

\bibitem{DBLP:conf/cascon/LiuXYCZ15}
H.~Liu, M.~Xu, Z.~Yu, V.~Corvinelli, and C.~Zuzarte.
\newblock Cardinality estimation using neural networks.
\newblock In {\em CASCON}, 2015.

\bibitem{lohmanblog}
G.~Lohman.
\newblock Is query optimization a solved problem?
\newblock \url{http://wp.sigmod.org/?p=1075}, 2014.

\bibitem{DBLP:conf/cidr/MalikBC07}
T.~Malik, R.~C. Burns, and N.~V. Chawla.
\newblock A black-box approach to query cardinality estimation.
\newblock In {\em CIDR}, pages 56--67, 2007.

\bibitem{DBLP:conf/sigmod/MarcusP18}
R.~Marcus and O.~Papaemmanouil.
\newblock Deep reinforcement learning for join order enumeration.
\newblock In {\em International Workshop on Exploiting Artificial Intelligence
  Techniques for Data Management}, 2018.

\bibitem{DBLP:journals/pvldb/MoerkotteNS09}
G.~Moerkotte, T.~Neumann, and G.~Steidl.
\newblock Preventing bad plans by bounding the impact of cardinality estimation
  errors.
\newblock {\em {PVLDB}}, 2(1):982--993, 2009.

\bibitem{DBLP:journals/pvldb/MullerMK18}
M.~M{\"{u}}ller, G.~Moerkotte, and O.~Kolb.
\newblock Improved selectivity estimation by combining knowledge from sampling
  and synopses.
\newblock {\em {PVLDB}}, 11(9):1016--1028, 2018.

\bibitem{DBLP:conf/sigmod/NeumannR18}
T.~Neumann and B.~Radke.
\newblock Adaptive optimization of very large join queries.
\newblock In {\em SIGMOD}, 2018.

\bibitem{DBLP:conf/sigmod/OrtizBGK18}
J.~Ortiz, M.~Balazinska, J.~Gehrke, and S.~S. Keerthi.
\newblock Learning state representations for query optimization with deep
  reinforcement learning.
\newblock In {\em Workshop on Data Management for End-To-End Machine Learning},
  2018.

\bibitem{DBLP:conf/vldb/PoosalaI97}
V.~Poosala and Y.~E. Ioannidis.
\newblock Selectivity estimation without the attribute value independence
  assumption.
\newblock In {\em VLDB}, 1997.

\bibitem{szegedy2016rethinking}
C.~Szegedy, V.~Vanhoucke, S.~Ioffe, J.~Shlens, and Z.~Wojna.
\newblock Rethinking the inception architecture for computer vision.
\newblock In {\em Proceedings of the IEEE conference on computer vision and
  pattern recognition}, pages 2818--2826, 2016.

\bibitem{DBLP:journals/pvldb/VengerovMZC15}
D.~Vengerov, A.~C. Menck, M.~Za{\"{\i}}t, and S.~Chakkappen.
\newblock Join size estimation subject to filter conditions.
\newblock {\em {PVLDB}}, 8(12):1530--1541, 2015.

\bibitem{DBLP:conf/sigmod/WuNS16}
W.~Wu, J.~F. Naughton, and H.~Singh.
\newblock Sampling-based query re-optimization.
\newblock In {\em SIGMOD}, pages 1721--1736, 2016.

\bibitem{zaheer2017deep}
M.~Zaheer, S.~Kottur, S.~Ravanbakhsh, B.~Poczos, R.~R. Salakhutdinov, and A.~J.
  Smola.
\newblock Deep sets.
\newblock In {\em Advances in Neural Information Processing Systems}, 2017.

\end{thebibliography}
